\documentclass[12pt,a4paper]{article}
\usepackage{amssymb}
\def\t#1{\tilde #1}

\def\b#1{{\mathbb #1}}

\def\c#1{{\cal #1}}

\def\Dirac{{\raise0.09em\hbox{/}}\kern-0.69em D}

\def\kbar {{\mathchar'26\mkern-9muk}}

\def\ep{i\epsilon}

\def\pprime{{\prime\prime}}
\def\tfrac #1#2{\textstyle{\frac{#1}{#2}}}
\def\dfrac #1#2{\displaystyle{\frac{#1}{#2}}}

\def\k{\kern-.1em\mathbin{,}\kern-.1em}
\def\hk{\kern.12em\raise-1em\hbox{$\hat{\raise1em\hbox{,}}$}\kern.12em}

\newcommand{\initiate}{\setcounter{equation}{0}}

\begin{document}

\title{A Dynamical 2-dimensional Fuzzy Space}

\author{M. Buri\'c$\strut^1$,   J. Madore$\strut^{2}$ \\
      $\strut^{1}$Faculty of Physics, P.O. Box  368\\
      11001 Belgrade
\and    $\strut^{2}$Laboratoire de Physique Th\'eorique\\
      Universit\'e de Paris-Sud, B\^atiment 211, F-91405 Orsay}

\date{}

\maketitle

\abstract{The noncommutative extension of a dynamical 2-dimen\-sional
space-time is given and some of its properties discussed. Wick
rotation to euclidean signature yields a surface which has as
commutative limit the doughnut but in a singular limit in which the
radius of the hole tends to zero.}

\vskip0.5cm


\parskip 4pt plus2pt minus2pt

\initiate

\section{Introduction and notation}

There is a very simple argument due to Pauli that the quantum effects
of a gravitational field will in general lead to an uncertainty in the
measurement of space coordinates. It is based on the observation that
two `points' on a quantized curved manifold can never be considered as
having a purely space-like separation. If indeed they had so in the
limit for infinite values of the Planck mass, then at finite values
they would acquire for `short time intervals' a time-like separation
because of the fluctuations of the light cone.  Since the `points' are
in fact a set of four coordinates, that is scalar fields, they would
not then commute as operators. This effect could be considered
important at least at distances of the order of Planck length, and
perhaps greater. This is one motivation to study noncommutative
geometry. A second motivation, which is the one we consider ours, is
the fact that it is possible to study noncommutative differential
geometry, and there is no reason to assume that even classically
coordinates commute at all length scales. One can consider for example
coordinates as order parameters as in solid-state physics and suppose
that singularities in the gravitational field become analogs of core
regions; one must go beyond the classical approximation to describe
them.  A straightforward and conservative way is to represent
coordinates by operators. The space-time manifold is thus replaced
with an algebra generated by a set of noncommutative `coordinates'.
The essential element which allows us to interpret a noncommutative
algebra as a space-time is the possibility~\cite{Wor89,Con94} to
introduce a differential structure on the former.

We define noncommutative `space' as an associative $*$-algebra $\c{A}$
generated by a set of hermitean `coordinates' $x^i$ which in some
limit tend to the (real) coordinates $\t{x}^i$ of a manifold; the
latter we identify as the classical limit of the geometry.  We 
suppose that the center of the algebra  $\c{A}$ is trivial.
The coordinates satisfy a set of commutation relations
\begin{equation}
[x^i,x^j] = i\kbar J^{ij}(x^k).
\end{equation}
The parameter $\kbar$ is introduced to describe the fundamental area
scale on which noncommutativity becomes important. It is presumably of
order of the Planck area $G\hbar$; the commutative limit is defined by
$\kbar\to 0$. We stress that we do not imply that all noncommutative
geometries are suitable as noncommutative models of space-time, any
more so than for example an exotic $\b{R}^4$ would be suitable as
commutative models.  

In order to define the differential structure on $\c{A}$ we use a
noncommutative version of Cartan's frame formalism~\cite{Mad00c}.  In
ordinary geometry a vector field can be defined as a derivation of the
algebra of smooth functions; this definition can be used also when the
algebra is noncommutative. A derivation, we recall, is a linear map
which satisfies the Leibniz rule; sometimes this is modified to a
`twisted' Leibniz rule, \cite{DimJomMolTsoWesWoh03,DimMue04}).  The
set of all derivations we denote by $\mbox{Der}(\c{A})$. The classical
notion of the moving frame is generalized in the following way.  The
frame on $\c{A}$ is a set of $n$ inner derivations $e_a$, generated by
`momenta' $p_a$:
\begin{equation}
e_af = [ p_a,f].                               \label{ep}
\end{equation}
We assume that the momenta also generate the whole algebra $\c{A}$.
For inner derivations, the Leibniz rule is the Jacobi identity.  An
alternative way to define the frame is to use the 1-forms $\theta^a $
dual to $e_a$ such that the relation
\begin{equation}
\theta^a(e_b) = \delta^a_b                          \label{jmmb}
\end{equation}
holds. The module of 1-forms we denote by $\Omega^1(\c{A}) $. To
define the left hand side of the equation (\ref{jmmb}), that is the
basic forms $\theta^a$, we first define the differential, exactly as
in the classical case, by the condition
\begin{equation}
df(e_a) = e_a f,                   \label{diff}
\end{equation}
and the  multiplication of 1-forms by elements of the algebra
$\c{A}$ by
\begin{equation}
f dg = f e_ag \theta^a  ,\qquad                      \label{bi1}
dg f = e_ag f \theta^a.
\end{equation}
Since every 1-form can be written as sum of such terms, the definition
of differential is complete. In particular, since
\begin{equation}
f \theta^a(e_b) = f\delta^a_b = (\theta^a  f)(e_b),   \label{cond}
\end{equation}
we conclude that the frame necessarily commutes with all the
elements of the algebra $\c{A}$.
The 1-form $\theta$ defined as
$
\theta = -p_a\theta^a
$
can be considered as an analog of the Dirac operator in ordinary
geometry.  It implements the action of the exterior derivative on
elements of the algebra, that is
$
df =  -\left[ \theta,f\right]
$.

The differential is real if $(df)^* = df^*$.  This is assured if the
derivations $e_a$ are real: $e_af^* = (e_a f)^* $, which is the case
if the momenta $p_a$ are antihermitean. 

Let us mention  further properties of the module structure
defined by~(\ref{diff}-\ref{cond}). The exterior product is 
a map from the tensor product of two copies of the module of
1-forms into the module of 2-forms; we shall identify the latter
as a subset of the former and write the product as
\begin{equation}
\theta^a\theta^b =
P^{ab}{}_{cd}\theta^c\otimes\theta^d.                   \label{exterior}
\end{equation}
The $ P^{ab}{}_{cd}$ are complex numbers which satisfy the projector
condition and the hermiticity~\cite{CerFioMad00a}:
\begin{equation}
P^{ab}{}_{cd}P^{cd}{}_{ef}=P^{ab}{}_{ef},\qquad 
\bar P^{ab}{}_{cd}P^{dc}{}_{ef}=P^{ba}{}_{ef} .           \label{pr}
\end{equation} 
The basis 1-forms anticommute for
$P^{ab}{}_{ cd} = \frac12\, (\delta^a_c\delta^b_d-\delta^b_c\delta^a_d)$.
The exterior derivative of $\theta^a$ is a 2-form, 
\begin{equation}
d\theta^a = -\frac 12 C^a{}_{\ bc}\, \theta^b\theta^c.         \label{s-e2}
\end{equation}
The $C^a{}_{ bc}$ are called the structure elements. 

The relation $d^2=0$ and the consistency of the relation (\ref{cond})
with the differential, $d(f\theta^a-\theta^a f)=0$, have nontrivial
consequences.  The structure elements are linear in the momenta
\begin{equation}
C^a{}_{bc}=F^a{}_{bc}-2p_dP^{(ad)}{}_{bc}   .                           \label{c}
\end{equation} 
Furthermore, the momenta obey a  quadratic relation
\begin{equation}
2 p_c p_d P^{cd}{}_{ab} -p_c  F^c{}_{ab} - K_{ab} = 0.   \label{quadratic}
\end{equation}
The $F^a{}_{bc}$ and $K_{ab}$ are complex numbers.
From (\ref{c}) it follows immediately that
\begin{equation}
e_a C^a{}_{bc} = 0. \label{con}
\end{equation}
This relation must be also satisfied in the commutative limit and
constitutes a constraint on the frame. A frame has four degrees of
freedom in two dimensions; the constraint subtracts one therefrom.

\initiate
\section{Fuzzy doughnut}   \label{doughnut}

Having outlined the main features of the frame formalism, let us
discuss it in a simple 2-dim case.  Clearly, every choice of a frame
$\theta^a$ implements a different differential structure. On the other
hand, the conditions (\ref{c}-\ref{quadratic}) constrain the possible
choices quite rigidely. This can easily be seen in low dimensions: one
can readily `solve' a family of 2-dim metrics with one Killing vector.
We shall exhibit all possible frames which yield differential calculi
based on inner derivations.  As a frame we choose
\begin{equation}
\theta^0 = f(x) dt, \quad f > 0, \qquad \theta^1 = dx.      \label{frame}
\end{equation}
  The frame relations can be  written as
\begin{equation}
\begin{array}{ll}
dx\, x = x\, dx, & \quad dx\, t = t\, dx, \\[4pt]
dt\, x =  x\, dt, &\quad dt\, t = \big(t + i\kbar F)dt,              \label{33}
\end{array}
\end{equation}
and imply  $dJ^{01}=0$. 
We have set
\begin{equation}
F = J^{01}\frac{d}{dx} \log f.
\end{equation}
The differential structure of the algebra can be written as
\begin{equation}
(dx)^2 = 0,\quad dx\, dt = - dt\, dx,\qquad
( dt)^2 = -\tfrac 12 i\kbar F^\prime dx\, dt
\end{equation}
or as the relations
\begin{eqnarray}
&&(\theta^1)^2 = 0,\quad \theta^0 \theta^1 = -\theta^1 \theta^0, \\[4pt]
&&(\theta^0)^2 = \tfrac 12 i\kbar  f F^\prime \theta^0 \theta^1
= 2\ep  \theta^0 \theta^1.
\end{eqnarray}
We have introduced here a parameter $\epsilon$ define by
\begin{equation}
\epsilon = \kbar \mu^2, \qquad \mu^2 = \tfrac 14 f F^\prime.
\end{equation}
It follows from the frame properties that the mass scale $\mu$ is a constant.

Suppose now that the dual momenta exist. The duality relations are
\begin{equation}
\begin{array}{ll}
[p_{0}, t] = f^{-1}, &\quad [p_{0}, x] = 0, \\[2pt]
[p_{1}, t] = 0, &\quad [p_{1}, x] = 1.
\end{array}
\end{equation}
If we denote $[p_{0}, p_{1}] = L_{01}$, the Jacobi
identities imply the relations
\begin{equation}
\begin{array}{ll}
[p_0,J^{01}] = 0,  &\quad [p_1,J^{01} ] =0,   \\[4pt]
[t,L_{01}] = -  f^\prime  f^{-2}, &\quad [x,L_{01}] =0.
\end{array}                                              \label{RF1}
\end{equation}
One can conclude again that $J^{01}$ is constant
and also that $L_{01}$ is a function of $x$ alone. We set
$J^{01}=1$. It follows that, neglecting the integration constants,
the `Fourier transformation' between the position and momentum
generators is given by
\begin{equation}
p_0 = - \dfrac{1}{i\kbar} \int f^{-1}, \qquad
p_1 = - \dfrac{1}{i\kbar}\, t.                         \label{tort}
\end{equation}
Each of the pairs $(t,x)$ and $(p_0, p_1)$ generates the algebra.

The array $P^{ab}{}_{cd}$ we write as
\begin{equation}
P^{ab}{}_{cd} =  \tfrac 12 \delta^{[a}_{c}\delta^{b]}_d +
\ep Q^{ab}{}_{cd} .
\end{equation}
In dimension two, if we assume that metric depends on $x$, that is
on $p_0$ only, we find that
\begin{equation}
P^{ab}{}_{cd} p_a p_b = \frac 12 [p_c, p_d] + \ep Q^{00}{}_{cd} p^2_0
\end{equation}
and therefore $L_{01}$ is given by
\begin{equation}
L_{01} = K_{01} + p_0  F^0{}_{01} - 2 \ep p_0^2 Q^{00}{}_{01}.
\end{equation}
The structure elements are 
\begin{equation}
C^0{}_{01} = F^0{}_{01} - 4 \ep p_0 Q^{00}{}_{01}.
\end{equation}
Symmetry and reality of the product (\ref{pr}) imply that $Q^{ab}{}_{cd}$ has 
the following non-vanishing elements:
\begin{equation}
Q^{10}{}_{00} = - Q^{01}{}_{00} = 1,\qquad
Q^{00}{}_{01} = - Q^{00}{}_{10} = 1.                \label{Qq}
\end{equation}
We set also
\begin{equation}
K_{01} = \frac 1{i\kbar J^{01}} = \frac 1{i\kbar}, \qquad 
F^0{}_{01} =  - ib\mu,
\end{equation}
while $C^0{}_{10}$ is determined by the constraint
\begin{equation}
C^0{}_{ab} P^{ab}{}_{01} = C^0{}_{01}, \qquad
C^0{}_{01} + C^0{}_{10} = -2\ep C^0{}_{00}.
\end{equation}
We have then finally the expressions
\begin{eqnarray}
&&L_{01} = (i\kbar)^{-1}(1 - b \mu^{-1}(\ep p_0) -
2\mu^{-2} (\ep p_0)^2), \\[8pt]
&&C^0{}_{01} = - ib \mu - 4\ep p_0,
\end{eqnarray}
and a differential equation for $p_0$:
\begin{equation}
-\ep \frac{dp_0}{dx} = \mu^2 - 
\ep b \mu p_0 - 2 (\ep p_0)^2    .                         \label{equa}
\end{equation}
 
There are three cases to be considered.  The simplest is the case with
$\mu^2 \to \infty$. The equation (\ref{equa}) reduces to
\begin{equation}
-i\kbar\,\dfrac{dp_0}{dx} = 1.                        \label{pp3}
\end{equation}
One finds the relations
\begin{equation}
i\kbar p_0 = - x, \qquad f(x) = 1.                   \label{S13}
\end{equation}
This is  noncommutative Minkowski space.

An equally degenerate case is the case $\mu^2 \to \infty$ and 
$\epsilon b=c\mu$. Equation~(\ref{equa}) can be written in the form
\begin{equation}
- i\kbar\, \dfrac{dp_0}{dx} = 1 - ic p_0.                \label{pp2}
\end{equation}
One finds the solution
\begin{equation}
i p_0 = c^{-1}(e^{-\kbar^{-1} cx} - 1), \qquad 
f(x) =  e^{\kbar^{-1} c x}.                            \label{S12}
\end{equation}
This  is noncommutative de~Sitter space; it can be brought to the usual
form by the change of variables
\begin{equation}
t^\prime = 2t, \qquad \mu x^\prime = 2c^{-1}(e^{-c x}-1).
\end{equation}

The case which interests us the most  is that with $\mu$ finite.
With $b=0$ (that is, with $F^a{}_{bc}=0$) the equation (\ref{equa}) becomes
\begin{equation}
-\ep\, \frac{dp_0}{dx} = \mu^2 -2 (\ep p_0)^2.
\end{equation}
If we fix $\beta^2 = 2 \mu^2 > 0$,
the equation for $p_0$ becomes
\begin{equation}
\frac{1}{\beta} \frac{d}{dx} \left(-2 \ep\beta^{-1} p_0 \right)
= 1 - \left(- 2\ep \beta^{-1} p_0 \right)^2.
\end{equation}
The solution to this equation is given by
\begin{equation}
i\kbar p_0 = - \beta^{-1} \tanh (\beta x),
\end{equation}
with 
 \begin{equation}
f(x) = \cosh^2 (\beta x)                    \label{S1}
\end{equation}
and
\begin{equation}
F = - 2i\beta^2 \kbar p_0 = 2\beta \tanh(\beta x).
\end{equation}
The frame corresponding to this solution is given by
\begin{equation}
\theta^0 = \cosh^2(\beta x)dt =
\frac 12 (1 + \cosh (2\beta x)) dt, \qquad
\theta^1 = dx .
\end{equation}
Frames of similar type have
appeared~\cite{LemSa94,GegKun97,GruKumVas02} in 2-dimensional dilaton
gravity.  The connection and the
curvature of the analogous commutative moving  frame are
\begin{eqnarray}
&&\t{\omega}^0{}_1 = \t{\omega}^1{}_0 = F \t{\theta}^0, \\[4pt]
&&\t{\Omega}^0{}_1 = \t{\Omega}^1{}_0 =
-(F^\prime +F^2)\theta^0\theta^1 = - f^{-1} f^\pprime
\t{\theta}^0\t{\theta}^1.                                \label{lcm}
\end{eqnarray}
The solution is a completely regular manifold of Minkowski signature.
In the limit $\beta\to 0$
\begin{equation}
i\kbar p_0 = - x, \qquad f = 1,
\end{equation}
 one finds Minkowski space.  In `tortoise' coordinate $x^*$,
$\ x^* = \int \dfrac{dx}{f(x)} ,\ $ 
the frame is given by
\begin{equation}
\theta^0 = \dfrac {1}{1-x^{*2}}\,  dt, \qquad
\theta^1 = \dfrac {1}{1-x^{*2}}\,  dx^*.
\end{equation}
From~(\ref{tort}) we see that $x^* = -i\kbar p_0$.

Under a Wick rotation
\begin{equation}
u=2i\beta x, \qquad v= t,
\end{equation}
the frame~(\ref{frame}) becomes
\begin{equation}
\theta^0 = \frac 12 (1 + \cos u) dv, \qquad
\theta^1 = \dfrac{1}{2i\beta}\,du ,
\end{equation}
and the corresponding commutative line element has the form
\begin{equation}
d\t{s}^2 = \frac 14 \,(1 +\cos\t{ u})^2  d\t{v}^2 + 
\frac {1}{4}\,\beta^{-2} d\t{u}^2.
\end{equation}
This is the surface of the torus embedded in $\b{R}^3$:
\begin{equation}
\t{ x}=\frac 12 (1+\cos \t{u})\cos\t{ v},\quad
\t{y}= \frac 12 (1+\cos \t{u})\sin\t{ v},\quad
\t{ z} =\frac 1{2} \beta^{-1}\sin \t{u},
\end{equation}
and for this reason we call this metric the `fuzzy doughnut'.  It is a
singular axially-symmetric surface of Gaussian curvature
\begin{equation}
\t{K} = 2\beta^2 (1 - \tan^2 \frac 12\, \t{u}).
\end{equation}
The doughnut is defined by the coordinate range $0 \leq \t{u} \leq 2\pi$,
$0 \leq \t{v} \leq 2\pi$, with a singularity at the point $\t{u} =\pi$.
In spite of the singularity, the Euler characteristic is given by
\begin{equation}
e[\c{A}] = \frac{1}{4\pi} \epsilon_{ab} \int \t{\Omega}^{ab} =
 - \frac{1}{2\pi} \int \t{\Omega}^0{}_1 =
 - \frac{1}{2\pi} \int d\t{\omega}^0{}_1 = 0
\end{equation}
as it should be.  If we suppose the same domain in the Wick rotated
real-$t$ region, then
\begin{equation}
0 \leq x \leq \beta^{-1}\pi, \qquad
0 \leq t \leq 2\pi .
\end{equation}
As $\beta\to \infty$ the doughnut becomes more and more squashed,
and this domain becomes an elementary domain in the limiting Minkowski
space.

\initiate
\section{Noncommutative differential geometry}             \label{dg}

In order to investigate the differential-geometric structure of the
fuzzy doughnut we mention first the definition the linear connection and
the metric, specified in the frame formalism; for details we refer
to~\cite{Mad00c,BurMacMad04}. Note that when the momenta exist the
metric is given; otherwise there is a certain ambiguity which must be
determined by field equations. Next, we will apply these definitions
in the weak-field approximation $\epsilon \to 0$, and find the first
noncommutative corrections to the classical doughnut geometry.

To define a linear connection one needs a
`flip'~\cite{Mou95,DubMadMasMou96b}, 
$
\sigma(\theta^a \otimes \theta^b) = S^{ab}{}_{cd} \theta^c \otimes \theta^d,
$
which in the present notation is
equivalent to a 4-index set of complex numbers $S^{ab}{}_{cd}$ which
we can write as
\begin{equation}
S^{ab}{}_{cd} = \delta_c^b \delta_d^a + \ep T^{ab}{}_{cd}.
\end{equation}
 The covariant derivative of a 1-form $\xi$  is given by
$
D \xi = \sigma(\xi \otimes \theta) - \theta \otimes \xi .
$
In particular
\begin{equation}
D \theta^a = -\omega^a{}_c\otimes\theta^c = - (S^{ab}{}_{cd} - 
\delta^b_c \delta^a_d) p_b \theta^c \otimes \theta^d = 
-\ep T^{ab}{}_{cd}  p_b \theta^c \otimes \theta^d ,
\end{equation}
so the connection-form coefficients are linear in the momenta
\begin{equation}
\omega^a{}_c = \omega^a{}_{bc}\theta^b = 
\ep p_d T^{ad}{}_{bc}\theta^b .                         \label{LeviC}
\end{equation}
On the left-hand side of the last equation is a quantity
$\omega^a{}_c$ which measures the variation of the metric; on the
right-hand side is the array $T^{ad}{}_{bc}$ which is directly related
to the anti-commutation rules for the 1-forms, and more important the
momenta $p_d$ which define the frame. As $\kbar \to 0$ the right-hand
side remains finite and
$
\omega^a{}_{c} \to \t{\omega}^a{}_{c}.
$
The identification is only valid in the weak-field approximation. 
The connection is torsion-free if the components satisfy the constraint
\begin{equation}
\omega^a{}_{ef}P^{ef}{}_{bc} = \tfrac 12 C^a{}_{bc}.          \label{t-free}
\end{equation}

The metric is a bilinear map
$
g:\; \Omega^1(\c{A}) \otimes \Omega^1(\c{A}) \rightarrow \c{A}.
$
It can be defined using the frame as
\begin{equation}
g(\theta^a\otimes\theta^b) = g^{ab}.
\end{equation}
The bilinearity of the metric implies that the $g^{ab}$ must belong to
the center of the algebra; they must be complex numbers. This is a familiar
property of the moving frame formalism.  Here we take the frame to be
orthonormal in the commutative limit and thus to first order the
metric is
\begin{equation}
g^{ab} = \eta^{ab} +\ep h^{ab}.
\end{equation}
In general it is unclear whether the noncommutative extension can
always be defined so that it possesses all usual properties.  In the
present formalism~\cite{Mad00c} the metric is `real' if it satisfies
the condition
$
\bar g^{ba} = S^{ab}{}_{cd}g^{cd}.               \label{metreal}
$
The definition of `symmetry' of the metric is ambiguous: it can be defined
either using the projection,
$
P^{ab}{}_{cd}g^{cd} = 0,                     \label{metsymP}
$
or the flip
$
S^{ab}{}_{cd}g^{cd} = cg^{ab}.             \label{metsymS}
$
We shall see that the present example prefers the former definition.
 
In general a connection is metric-compatible if the condition
\begin{equation}
\omega^i{}_{kl} g^{lj} +
\omega^j{}_{ln} S^{il}{}_{km} g^{mn} = 0                     \label{metcom-c}
\end{equation}
is satisfied. This can be written in a more familiar form
$
D_i g^{jk} = 0
$
if one introduce~\cite{Mad00c} an appropriately twisted `covariant derivative'.
Linearized, the condition reads
\begin{equation}
T^{(ac}{}_d{}^{b)} = 0.                      \label{complin}
\end{equation}
One should note that not all of the usual conditions on a metric are
necessarily satisfied. It is therefore of interest to see, in specific
examples, which of them can be imposed. Although the relations are
algebraic, they form a set of equations difficult to solve in full
generality even in dimension two. They simplify somewhat if one use
a perturbation expansion.

In our 2-dim model the frame is of the form
\begin{equation}
\theta^0 = f(x) dt, \qquad
\theta^1 = dx.
\end{equation}
The torsion-free metric-compatible connection and the curvature are
classically given by the expressions~(\ref{lcm}). From these
expressions we see that the geometry is flat only if $f(x)$ is linear
in $x$. We recall that $\epsilon = \kbar\mu ^2$. To first order the
fuzzy calculus differs from the commutative limit in the two relations
\begin{equation}
\theta^{(0} \theta^{1)} = -2 \ep q  (\theta^0)^2, \quad
(\theta^0)^2  = \ep q  \theta^{[0} \theta^{1]},                               
\end{equation}
which to first order reduce to 
\begin{equation}
\theta^{(0} \theta^{1)} = 0, \quad
(\theta^0)^2  = 2 \ep q  \theta^{0} \theta^{1}.                  \label{20} 
\end{equation}
The quantity $q$ which we have introduced in (\ref{20}) is a
constant: $q =0$ in the cases of flat and of de~Sitter noncommutative
space, $q=1$ in the fuzzy doughnut case. We will restrict our
considerations to the latter.

The differentials of the frame are given by
\begin{equation}
d\theta^0 = - C^0{}_{01} \theta^0 \theta^1, \qquad d\theta^1 = 0,
\end{equation}
with
$
C^0{}_{01} = - 4 \ep p_0 Q^{00}{}_{01} = - 4\ep p_0.
$
The only non-vanishing components of the connection are
\begin{equation}
\omega^0{}_1 = \omega^1{}_0 = -4\ep p_0 \theta^0 = F \theta^0,
\end{equation}
so from (\ref{LeviC}) we  find 
\begin{equation}
T^{00}{}_{01} = T^{10}{}_{00} = -4.                     \label{t-us}
\end{equation}
Metric-compatibility (\ref{complin}) is satisfied to first order. Also, to
the same order the condition that the torsion vanish~(\ref{t-free}) is
 satisfied by the values we obtain.
The curvature 2-form has components
\begin{equation}
\Omega^0{}_1 = - (F^\prime + F^2) \,\theta^0 \theta^1,\qquad
\Omega^0{}_0 = \Omega^1{}_1= 2\ep F^2 \theta^0 \theta^1.
\end{equation}

We must define a `real', `symmetric' metric.  There are in principle
four possible ways to define it depending on which of two possible
ways one chooses to define symmetry, and whether or not one includes a
twist in the extension of the metric to the tensor product.  
In the present example one sees that the only consistent choice is the 
following 
\begin{equation}
h^{[ab]} = - 2 Q_+^{abcd} \eta_{cd} = - 2 Q_-^{cdab} \eta_{cd},
\end{equation}
where
 we denoted $Q_-^{ab}{}_{cd} = \frac 12 Q^{ab}{}_{[cd]}$,  
$Q_+^{ab}{}_{cd} = \frac 12 Q^{ab}{}_{(cd)}$.
Thus for the symmetric and real  metric we obtain
\begin{equation}
g^{ab} = \eta^{ab} + \ep h^{ab} = 
\left(\begin{array}{cc} -1 &2\ep \\[2pt] 0 &1 \end{array}\right).
\end{equation}

\initiate
\section{Higher-order effects}     \label{ho} 

To find the second order corrections to our system, we write the
4-index tensors as matrices ordering the indices $(01,10,11,00)$.  Let
$P_0$ and $S_0$ be respectively the canonical projector and the flip
\begin{equation}
P_0= \frac 12 \left(\begin{array}{cccc}
 1 & -1 & 0 & 0\\
-1 &  1 & 0 & 0\\
 0 &  0 & 0 & 0\\
 0 &  0 & 0 & 0
\end{array}\right) ,\qquad
S_0=\left(\begin{array}{cccc}
0 & 1 & 0 & 0\\
1 & 0 & 0 & 0\\
0 & 0 & 1 & 0\\
0 & 0 & 0 & 1
\end{array}\right) ,
\end{equation} 
and denote the second order projection and flip by $P$ and $S$
\begin{eqnarray}
&&P = P_0 + \ep Q + (\ep)^2 Q_2,  \\[4pt]
&&S = S_0 + \ep T + (\ep)^2 T_2 .
\end{eqnarray}
The projector constraints are, in matrix notation,
\begin{equation}
P^2 = P,\qquad   \bar  P \hat P = \hat P ,                 \label{p2p}
\end{equation}
where $\hat A^{ab}{}_{cd} = A^{ba}{}_{cd} = (S_0 A)^{ab}{}_{cd}.$ 
To first order the projector conditions become
\begin{equation}
\bar Q = Q, \qquad \hat Q_- = Q_-   .                  
\end{equation}
The twist constraints are
\begin{equation}
\hat {\bar S} \hat S = 1, \qquad
\hat S P + \bar P \hat P = 0, \qquad
SP + P = 0 .                                                                                 \label{S}         
\end{equation}
The last two identities are equivalent if
$
\bar P \hat P = \hat P,
$
the condition already imposed in (\ref{p2p}).

One can easily check  the first order solution of the previous
section. In the matrix notation it is
given by
\begin{equation}
Q_- = \left(\begin{array}{cc}
0  &0 \\
\tau  &0
\end{array}\right)   ,      \quad
Q_+ = \left(\begin{array}{cc}
0  &-\tau^* \\
0  &0
\end{array}\right),  \quad  Q = Q_- + Q_+ =\left(
\begin{array}{cccc} 
0& 0& 0& -1\\
0& 0& 0& 1\\
0& 0& 0& 0\\
1& -1& 0& 0
\end{array}\right) ,
\end{equation} 
and 
\begin{equation}
T = - 2 \left(\begin{array}{cc}
0 & \tau\tau^*\\[2pt]
 \tau\tau^*\sigma_1 &0 \end{array}\right)  
=\left(
\begin{array}{cccc} 
0& 0& 0& 0\\
0& 0& 0& -4\\
0& 0& 0& 0\\
-4& 0& 0& 0
\end{array}\right)     ,             
\end{equation}
where we introduced the matrix $\tau$ and its transpose $\tau^*$, $\tau = \left(\begin{array}{cc}
0 &0\\
1 &-1\end{array}\right)$.

The constraints (\ref{p2p}-\ref{S}) can be solved to
second order using inner automorphisms of the matrix algebra.
Write $\ P = W^{-1} P_0 W\  $,
with $W$  arbitrary nonsingular $4\times 4$ matrix. We see immediately that
$P^2 = P $.
To satisfy the second condition of~(\ref{p2p}) on $P$ it is
sufficient to require that
\begin{equation}
\bar W S_0 = S_0 W     ,                           \label{barQS}
\end{equation}
and to recall  $\hat S = S_0 S$. Let
$W = \exp(\ep B)$.
To second order 
\begin{equation}
P = P_0 + \ep [P_0, B] + \tfrac 12 (\ep)^2 [[P_0,B],B],
\end{equation}
so the two expansions coincide if
\begin{equation}
Q = [P_0, B],     \qquad  Q_2 = \tfrac 12 [[P_0,B],B] = \tfrac 12 [Q,B].                \label{QmQp}
\end{equation}
It is easy to find the appropriate solution 
\begin{equation}
B  = \left(\begin{array}{cc}
0 &-\tau^* \\[2pt] 
-\tau  &0
\end{array}\right).
\end{equation}

The solution for $T_2$ is
\begin{equation}
 T_2 = \tfrac 12 T S_0 T. 
\end{equation}
To summarize,
\begin{equation}
P = \left(\begin{array}{cccc}
 1/2 -\epsilon^2& -1/2+\epsilon^2& 0& -i\epsilon\\
 -1/2+\epsilon^2 & 1/2-\epsilon^2&0& i\epsilon\\
 0& 0& 0&0\\
i\epsilon & -i\epsilon& 0&2\epsilon^2
\end{array}\right),
\qquad
S= \left(\begin{array}{cccc}
0& 1& 0 & 0\\
 1-8\epsilon^2& 0&0& -4i\epsilon\\
 0& 0& 1&0 \\
-4i\epsilon& 0& 0&1-8\epsilon^2
\end{array}\right).
\end{equation}

The second-order metric is given by
\begin{equation}
g = g_0 - \tfrac 12  \ep S_0 T g_0 =  = \left(\begin{array}{cc}
-1 &0 \\[2pt] 
0  &1
\end{array}\right) + 2\ep \left(\begin{array}{cc}
0 &1 \\[2pt] 
0  &0
\end{array}\right).
\end{equation}
It is real and symmetric.
We saw in Section~\ref{dg} that this metric is compatible with the
connection 
\begin{equation}
\omega^a{}_b = S^{ac}{}_{db}\theta^d p_c + \theta \delta^a_b, \qquad 
\theta = - p_a \theta^a
\end{equation}
to first order in the expansion parameter.  We have not succeeded in
finding a connection which is metric-compatible to second order. The
non-metricity was found in noncommutative spaces defined by some
string-theory models before~\cite{Saz03}.

\initiate
\section{Further speculations}

Several models have been found which illustrate a close relation
between noncommutative geometry in its `frame-formalism' version and
classical gravity. Heuristically, but incorrectly, one can formulate
the relation by stating that gravity is the field which appears when
one quantizes the coordinates much as the Schr\"odinger wave function
encodes the uncertainty resulting from the quantization of phase
space.

The first and simplest of these is the fuzzy sphere~\cite{Mad92a}
which is a noncommutative geometry which can be identified with the
2-dimensional (euclidean) `gravity' of the 2-sphere. The algebra in
this case is an $n\times n$ matrix algebra; if the sphere has radius
$r$ then the parameter $r/n$ can be interpreted as a lattice length.
With the identification this model illustrates how gravity can act as
an ultraviolet cutoff, a regularization which is very similar to the
`point splitting' technique which has been used when quantizing a
field in classical curved backgrounds. It can also be compared with
the screening of electrons in plasma physics, which gives rise to a
Debye length proportional to the inverse of the electron-number
density $n$. The analogous `screening' of an electron by virtual
electron-positron pairs is responsible for the reduction of the
electron self-energy from a linear to logarithmic dependence on the
classical electron radius.  We refer to a noncommutative geometry as
`fuzzy' if the algebra and the representation are such so that there
is a true `screening of points'. A notable counterexample would be the
irrational noncommutative torus.  Other models have been found which
illustrate the identification including an infinite series in all
dimensions.

In the present paper yet another model is given, one which although
representing a classical manifold of dimension 2 is of interest
because the classical `gravity' which arises has a varying Gaussian
curvature. The authors will leave to a subsequent article the delicate
task of explaining exactly which property of the metric makes it
`quantizable'.  This geometry could furnish a convenient model to
study noncommutative effects, for example in the colliding-$D$-brane
description of the Big-Bang proposed by Turok \& Steinhardt
\cite{TurSte04}. The 2-space describing the time evolution of the
separation of the branes has been shown to be conveniently described
using Rindler coordinates. One can blur this geometry by using the
metric and connection described here.  The flat geometry would have to
be replaced by the one given in this section; in the limit $q\to 0$ it
would become flat.

The doughnut example is of importance in that is is the first explicit
construction of an algebra and differential calculus which is
singularity-free in the Minkowski-signature domain and which has a
non-constant curvature. There are two aspects of this problem. To
construct a classical manifold from a differential calculus is
relatively simple once one has constructed the frame.  One takes
formally the limit and uses the so constructed moving frame to define
the metric. This is contained in the upper right of the following
little diagram
\begin{equation}
\begin{array}{ccc}
\buildrel{\mbox{Fuzzy}}\over{\mbox{Frame}}&\longrightarrow
&\buildrel{\mbox{Classical}}\over{\mbox{Frame}}\\[6pt]
\vert &&\vert\\[-5pt]
\downarrow &&\downarrow\\[6pt]
\buildrel{\mbox{\it Fuzzy}}\over{\mbox{\it Geometry}}&\longrightarrow
&\buildrel{\mbox{Classical}}\over{\mbox{Geometry}}
\end{array}
\end{equation}
More difficult is the construction of a `fuzzy geometry' which would
fill in the lower left of the diagram and would be such that the
classical geometry is a limit thereof. But this step is very important
since it gives an extension of the right-hand side into what could
eventually be a domain of quantum geometry. It is the box in the
to-be-constructed lower left corner where possibly one can find an
interesting extension of the metric containing correction terms which
describe the noncommutative structure.

We have not succeeded however to completely extend this geometry to
all orders in the noncommutativity parameter $\ep$. This will be
considered in a subsequent article. There is evidence that the
extension will involve a non-vanishing value of the torsion 2-form.
The metric is extended into the noncommutative domain so as to
maintain such formal properties as reality and symmetry. The
interpretation however as a length requires more attention when the
`coordinates' do not commute.

Last, but not least, our example illustrates even better than the
fuzzy sphere the way in which quantum mechanics is modified by
geometry and the important role which {\em noncommutative} geometry
plays in understanding the relation between the two. The `momenta'
which we introduce are the natural curved-space generalization of the
canonical momentum operators of ordinary quantum mechanics. In the
present formalism they generate the algebra as well as do the
coordinates. Once the algebra is given the noncommutative structure of
space-time is manifest in the commutation relations $[x^i,x^j]$ and
the appropriate curved-space version of quantum mechanics is defined
by the relations~(\ref{ep}). The two structures are intimately
enmeshed by the Fourier transform as well as Jacobi identities. If the
right-hand side of~(\ref{ep}) reduces to the Kronecker symbol when
$f=x^i$ then the space is flat; because of the Jacobi identities only in
this case can quantum mechanics be consistent with a commutative
space-time structure.

\section*{Acknowledgment}

Part of this work was done while the authors were visiting ESI in
Vienna.  They would like to thank H.~Grosse, as well as M.~Axenides,
M.~Floratos, T.~Grammatikopoulos, J.~Mourad, T.~Sch\"ucker and
G.~Zoupanos for enlightening conversations. The research was supported
in part by the Grant No. 1468 of the Serbian Ministry of Science.


\begin{thebibliography}{10}

\bibitem{Wor89}
S.~L. Woronowicz, ``Differential calculus on compact matrix pseudogroups,''
  {\em Commun.\ Math.\ Phys.} {\bf 122} (1989) 125.

\bibitem{Con94}
A.~Connes, {\em Noncommutative Geometry}.
\newblock Academic Press, 1994.

\bibitem{Mad00c}
J.~Madore, {\em An Introduction to Noncommutative Differential Geometry and its
  Physical Applications}.
\newblock No.~257 in London Mathematical Society Lecture Note Series. Cambridge
  University Press, second~ed., 2000.
\newblock 2nd revised printing.

\bibitem{DimJomMolTsoWesWoh03}
M.~Dimitrijevic, L.~Jonke, L.~Moller, E.~Tsouchnika, J.~Wess, and
  M.~Wohlgenannt, ``Deformed field theory on $\kappa$-spacetime,'' {\em Euro.\
  Phys.\ Jour.~C} {\bf C31} (2003) 129--138,
\href{http://xxx.lanl.gov/abs/hep-th/0307149}{{\tt hep-th/0307149}}.

\bibitem{DimMue04}
A.~Dimakis and F.~Mueller-Hoissen, ``Automorphisms of associative algebras and
  noncommutative geometry,'' {\em J.~Phys.~A:\ Math.\ Gen.} {\bf 37} (2004)
  2307--2330,
\href{http://xxx.lanl.gov/abs/math-ph/0306058}{{\tt math-ph/0306058}}.

\bibitem{CerFioMad00a}
B.~L. Cerchiai, G.~Fiore, and J.~J. Madore, ``Geometrical tools for quantum
  euclidean spaces,'' {\em Commun.\ Math.\ Phys.} {\bf 217} (2001), no.~3,
  521--554,
\href{http://xxx.lanl.gov/abs/math.QA/0002007}{{\tt math.QA/0002007}}.

\bibitem{LemSa94}
J.~P.~S. Lemos and P.~M. Sa, ``The black holes of a general two-dimensional
  dilaton gravity theory,'' {\em Phys.\ Rev.} {\bf D49} (1994) 2897--2908,
\href{http://xxx.lanl.gov/abs/gr-qc/9311008}{{\tt gr-qc/9311008}}.

\bibitem{GegKun97}
J.~Gegenberg and G.~Kunstatter, ``Solitons and black holes,'' {\em Phys.\
  Lett.} {\bf B413} (1997) 274--280,
\href{http://xxx.lanl.gov/abs/hep-th/9707181}{{\tt hep-th/9707181}}.

\bibitem{GruKumVas02}
D.~Grumiller, W.~Kummer, and D.~V. Vassilevich, ``Dilaton gravity in two
  dimensions,'' {\em Phys.\ Rep.} {\bf 369} (2002) 327--430,
\href{http://xxx.lanl.gov/abs/hep-th/0204253}{{\tt hep-th/0204253}}.

\bibitem{BurMacMad04}
M.~Buri{\'c}, M.~Maceda, and J.~Madore, ``On the resolution of space-time
  singularities {III},'' in {\em Geometric Methods In Physics}, A.~Odzijewicz,
  A.~Strasburger, S.~T. Ali, J.-P. Antoine, T.~Friedrich, J.-P. Gazeau,
  Z.~Hasiewicz, and M.~Schlichenmaier, eds., pp.~--.
\newblock 2004.
\newblock
Bialowieza, Poland, July 2004.

\bibitem{Mou95}
J.~Mourad, ``Linear connections in non-commutative geometry,'' {\em Class.\ and
  Quant.\ Grav.} {\bf 12} (1995) 965.

\bibitem{DubMadMasMou96b}
M.~Dubois-Violette, J.~Madore, T.~Masson, and J.~Mourad, ``On curvature in
  noncommutative geometry,'' {\em J.~Math.\ Phys.} {\bf 37} (1996), no.~8,
  4089--4102, \href{http://xxx.lanl.gov/abs/q-alg/9512004}{{\tt
  q-alg/9512004}}.

\bibitem{Saz03}
B.~Sazdovi\' c, ``Torsion and nonmetricity in the stringy geometry,''
\href{http://xxx.lanl.gov/abs/hep-th/0304086}{{\tt hep-th/0304086}}.

\bibitem{Mad92a}
J.~Madore, ``The fuzzy sphere,'' {\em Class. Quant. Grav.} {\bf 9} (1992)
69--88.

\bibitem{TurSte04}
N.~Turok and P.~J. Steinhardt, ``Beyond inflation: A cyclic universe
  scenario,'' {\em Physica Scripta} (2004)
\href{http://xxx.lanl.gov/abs/hep-th/0403020}{{\tt hep-th/0403020}}.

\end{thebibliography}

\providecommand{\href}[2]{#2}\begingroup\raggedright\endgroup

\end{document}